\documentclass[11pt]{abpaper}

\usepackage{mathpazo} 
\usepackage[scaled]{helvet} 
\usepackage{courier} 
\normalfont
\usepackage[T1]{fontenc}

\usepackage{mathtools}
\usepackage{xspace,url}
\usepackage{caption,graphicx}
\usepackage{siunitx}
\usepackage{booktabs}
\usepackage{relsize}
\usepackage{underscore}
\usepackage{hyperref}
\usepackage{pygtex}

\usepackage{todonotes}
\presetkeys{todonotes}{inline}{}

\def\kbdstyle{\bfseries\ttfamily}
\newcommand{\kbd}[1]{{\kbdstyle #1}\xspace}

\newcommand{\Cxx}{C\!\,\raisebox{0.05\baselineskip}{++}\xspace}
\newcommand{\Fortran}{\textsc{Fortran}\xspace}

\newcommand{\GeV}{\si{\giga\electronvolt}}

\newcommand{\Doc}{\kbd{Doc}}
\newcommand{\Block}{\kbd{Block}}
\newcommand{\Particle}{\kbd{Particle}}
\newcommand{\Decay}{\kbd{Decay}}
\newcommand{\Process}{\kbd{Process}}
\newcommand{\XSec}{\kbd{XSec}}

\renewcommand\linebreak[1][]{}

\author{Andy Buckley\\
  \smaller \textit{School of Physics \& Astronomy, University of Glasgow, UK}%
  \thanks{At \textit{School of Physics \& Astronomy, University of Edinburgh} and \textit{PH Dept, CERN} for first submission.}}
\title{PySLHA: a Pythonic interface to SUSY~Les~Houches~Accord data}
\preprints{MCnet-2013-006\\ GLAS-PPE/2015-02}

\begin{document}

\begin{abstract}
  This paper describes the PySLHA package, a Python language module and program
  collection for reading, writing and visualising SUSY model data in the SLHA
  format. PySLHA can read and write SLHA data in a very general way, including
  the official SLHA2 extension and user customisations, and with arbitrarily
  deep indexing of data block entries and a dedicated, intuitive interface for
  particle data and decay information. The draft SLHA3 \kbd{XSECTION} feature is
  also fully supported. PySLHA can additionally read and write the legacy ISAWIG
  model format, and provides format conversion scripts. A publication-quality
  mass spectrum \& decay chain plotting tool, \kbd{slhaplot}, is included in the
  package.
\end{abstract}


\section{Introduction}

The SUSY Les~Houches Accord (SLHA) data format\,\cite{Skands:2003cj} has proven
an effective and exceedingly popular mechanism for exchange of weak-scale
supersymmetry model information between spectrum generator codes
(e.g. SoftSUSY\,\cite{Allanach:2001kg}, SPheno\,\cite{Porod:2003um}, and many
others) to Monte Carlo event generators and other users of weak scale SUSY
parameters. As such, SLHA has become the current \emph{de facto} format for
exchange of SUSY model data, replacing various code-specific formats such as the\linebreak[4]
ISAWIG format\,\cite{isawig} used to exchange
ISAJET-generated\,\cite{Paige:2003mg} model data to the \Fortran HERWIG event
generator\,\cite{Corcella:2000bw}.

SLHA's text-based format was designed to be easily written and read by hand, and
by the prevalent \Fortran-based spectrum generators and MC codes at the time of
its definition. The original format was defined during and following the
Les~Houches 2003 workshop\,\cite{Skands:2003cj}, and was extended in 2007 to the
current version~2 of the SLHA standard\,\cite{Allanach:2008qq,slhaweb}. The
format consists of \emph{blocks} of model data, each identified by a header line
starting with either \kbd{BLOCK} or \kbd{DECAY} for model data blocks and
particle decays respectively. Each data block has a name and an associated
energy scale at which it is defined, as well as many \emph{entries}. Decay
blocks each define the total decay width and a list of constituent decay
channels \& branching ratios for a particle species (identified using the PDG MC
numbering scheme). The latest update of the SLHA standard introduces a
\kbd{XSECTION} block type for representation of calculated process
cross-sections at various energies and in different calculation schemes.

Several SLHA reader codes already exist in the \Fortran and \Cxx languages, but
the popular Python language was until recently unblessed with such
functionality. Python is an interpreted language, particularly well suited to
clear and effective expression of complex logic, and particularly has advantages
over \Fortran, \Cxx and other numerical/system programming languages where
``high-level'' operations such as rich string parsing are involved. This is
often the case in SUSY model exploration, where Python can act as a glue to tie
together the input and output stages of several distinct modelling programs. The
PySLHA library described in this note is a Python language module (and several
Python programs) for reading, writing, and converting SLHA model data, as well
as providing a high quality plotter of SUSY model mass spectra and decay chains,
hence providing a convenient and powerful integration of SLHA into the Python
``glue layer'' for BSM model studies.

\section{Library features}

PySLHA's main programmatic features are the \Doc,\linebreak[4] \Block, \Particle \&
\Decay, and \Process \& \XSec classes, which map directly on to features of the
SLHA data format. We first describe these structures and then the functions
which operate on them.

\subsection{\Doc}

From PySLHA 3.0.0, the main reading and writing functions in PySLHA accept and
return a single \Doc object; before this, they passed several \kbd{dict}s
of data objects, but the single \Doc is more robust and
extensible. \Doc is a simple structured container which provides access to
the main groups of SLHA data types, via its \kbd{blocks}, \kbd{decays} and
\kbd{xsections} attributes. A \kbd{write} method bound to the
\Doc allows convenient file and string I/O of its data, with controllable
precision, in addition to the unbound \kbd{write*} functions
described in Section~\ref{sec:io}.

A SLHA document can be accessed as follows:
%
\begin{Verbatim}[commandchars=\\\{\},frame=leftline,framesep=1.5ex,framerule=0.8pt,fontsize=\smaller]
\PY{o}{\PYZgt{}\PYZgt{}}\PY{o}{\PYZgt{}} \PY{k+kn}{import} \PY{n+nn}{pyslha}
\PY{o}{\PYZgt{}\PYZgt{}}\PY{o}{\PYZgt{}} \PY{n}{d} \PY{o}{=} \PY{n}{pyslha}\PY{o}{.}\PY{n}{read}\PY{p}{(}\PY{l+s}{\PYZdq{}}\PY{l+s}{myspectrum.slha}\PY{l+s}{\PYZdq{}}\PY{p}{)}
\PY{o}{\PYZgt{}\PYZgt{}}\PY{o}{\PYZgt{}} \PY{k}{print} \PY{n}{d}
\PY{o}{\PYZlt{}}\PY{n}{PySLHA} \PY{n}{Doc}\PY{p}{:} \PY{l+m+mi}{16} \PY{n}{blocks}\PY{p}{,} \PY{l+m+mi}{32} \PY{n}{decays}\PY{p}{,} \PY{l+m+mi}{94} \PY{n}{xsections}\PY{o}{\PYZgt{}}
\end{Verbatim}

In future versions, the \Doc (and the following types) will allow storage
of SLHA document inline and block comments as well as the parsed data. For now,
only non-comment data will be preserved in a read/write cycle.

If it can be imported, the \kbd{OrderedDict} Python type will be used as a base
type for the \kbd{blocks}, \kbd{decays} and \kbd{xsections} containers, which
have a \kbd{dict}-like interface, in order to ensure that the user-specified
ordering of data sections in the SLHA file is not disrupted by manipulation in
PySLHA.

\subsection{\Block}

From PySLHA 2.0.0 onward, the \Block class provides a similar interface to
Python's native \kbd{dict} type, i.e. an instance of \Block contains several
entries accessed by an index or key, either by named functions or via the
square-brackets \kbd{[$x$]} operator. \Block{}s also have an associated name and
a $Q$ value indicating the scale at which the contents are defined.

Like \kbd{dict}{}s, \Block instances may store any type of object as the value
of an entry, although the SLHA file format means that in practice this is
restricted to integers, floating point numbers, and character strings (and
sequences of the above). Also like \kbd{dict}{}s, \Block{}s support the Python
iterator protocol, the get/set-item syntax, and methods such as \kbd{items()},
\kbd{values()}, \kbd{keys()}, and \kbd{has_key()}. \emph{Unlike} \kbd{dict}{}s,
\Block entries can only be indexed on integers, or on tuples of integers -- or,
in a special case designed for handling the standard \kbd{ALPHA} block, a block
may contain a single unkeyed entry. The ability to use tuples of integers as
keys is particularly useful for blocks which represent mixing matrices, e.g.
%

\begin{Verbatim}[commandchars=\\\{\},frame=leftline,framesep=1.5ex,framerule=0.8pt,fontsize=\smaller]
\PY{n}{umix} \PY{o}{=} \PY{n}{pyslha}\PY{o}{.}\PY{n}{Block}\PY{p}{(}\PY{l+s}{\PYZdq{}}\PY{l+s}{UMIX}\PY{l+s}{\PYZdq{}}\PY{p}{,} \PY{n}{q}\PY{o}{=}\PY{l+m+mf}{1e7}\PY{p}{)}
\PY{n}{umix}\PY{p}{[}\PY{l+m+mi}{1}\PY{p}{,}\PY{l+m+mi}{1}\PY{p}{]} \PY{o}{=} \PY{l+m+mf}{0.1}
\PY{o}{.}\PY{o}{.}\PY{o}{.}
\PY{k}{print} \PY{n}{umix}\PY{p}{[}\PY{l+m+mi}{1}\PY{p}{,}\PY{l+m+mi}{1}\PY{p}{]}

\PY{n}{mymix} \PY{o}{=} \PY{n}{pyslha}\PY{o}{.}\PY{n}{Block}\PY{p}{(}\PY{l+s}{\PYZdq{}}\PY{l+s}{MYMIX}\PY{l+s}{\PYZdq{}}\PY{p}{)}
\PY{n}{mymix}\PY{p}{[}\PY{l+m+mi}{2}\PY{p}{,}\PY{l+m+mi}{1}\PY{p}{,}\PY{l+m+mi}{3}\PY{p}{]} \PY{o}{=} \PY{l+m+mf}{3.142}
\PY{o}{.}\PY{o}{.}\PY{o}{.}
\PY{k}{print} \PY{n}{mymix}\PY{p}{[}\PY{l+m+mi}{2}\PY{p}{,}\PY{l+m+mi}{1}\PY{p}{,}\PY{l+m+mi}{3}\PY{p}{]}
\end{Verbatim}

Adding and accessing multi-value block entries by integer (tuple) keys is
unambiguous when done programatically as above. PySLHA of course also has to be
able to parse SLHA entry lines into blocks and here some heuristics are
necessary to differentiate between keys and values. The approach adopted, via
the \kbd{add_entry} and \kbd{set_value} methods, is that if the supplied
argument is a string, it will be split into a tuple of strings; these are then
automatically converted to numeric types if possible, and consecutive strings
are combined together (to allow for string values containing spaces.) The
resulting tuple of ints, floats and strings is then parsed from left to right to
find the first non-int item: all items before this are treated as the key.

A strength of Python as a language for SLHA block parsing is that the language
is very dynamically typed: a variable can hold any type, and \kbd{dict}{}s,
\kbd{tuple}{}s, and other containers can hold heterogeneous entries. This is
what makes it possible for blocks to be indexed by integer tuples of any length
(or no length) and for values to either be scalar or tuples of mixed
types. Hence no special features are needed for SLHA2 support or further
extensions to the SLHA standard block content: any block layout representable in
the SLHA syntax (and some which are not) can be manipulated using the PySLHA
\Block class.

\subsection{\Particle and \Decay}

The \Particle and \Decay classes are the complements of \Block for SUSY particle
properties and decay specifications. Particles are defined by a PDG species
identifier code\,\cite{montecarlorpp} and a list of \Decay{}s, and optionally
may also contain the total decay width and the mass\footnote{There is
  intentional duplication here: the SLHA format design places particle masses in
  the \kbd{MASS} block, 
  but from a code design point of view it makes sense for a \Particle to know
  its own mass.} in \GeV. In this sense, \Particle{}s are representations of the
SLHA \kbd{DECAY} block itself, while its entries are each transformed into a
fully-fledged \Decay object. Each \Decay contains a branching ratio and a list
of particle ID codes representing the decay daughters. For
convenience and familiarity, the PDG ID of the decaying particle and the \kbd{NDA}
(number of daughters) number may also be stored but these are not essential: the
parent particle has its own fully fledged \Particle instance and \kbd{decay.nda}
is equivalent to the ``more Pythonic'' \kbd{len(decay.ids)}.

\subsection{\Process and \XSec}

The SLHA3 draft
standard\footnote{\url{https://phystev.cnrs.fr/wiki/2013:groups:tools:slha}}
introduces a new block type, the \kbd{XSECTION}, for storing various calculated
cross-sections for different processes. This is supported in PySLHA from version
3.1.0 onwards, via the \Process and \XSec types. As for the \kbd{DECAY} block, a
direct representation of the SLHA text format in memory is not the most natural
programming interface, and hence a slightly different structure is used.

The \Process class is the closest match to an \kbd{XSECTION} block, and the
\XSec objects contained within it map to the entries in the block. The main
distinction is that a different \kbd{XSECTION} block is used for each energy at
which a process cross-section is provided, which the \Process is a
\kbd{dict}-like object keyed on the concatenated tuple of (sorted) initial- and
final-state particle ID codes. A given \Process can then contain \XSec entries
at several energies, as well as the variations in QCD and EW order,
factorization and renormalization scale-factors, PDFs, and the computational
code which made the calculation. This is both more semantically natural, and
avoids the problem of indexing the dictionary on $\sqrt{s}$, a floating-point
number which could easily fail to compare exactly equal to the required key value.

The \XSec objects inside a \Process are available in the \kbd{xsecs} list
attribute. To make look-up of cross-sections which meet certain criteria easier,
however, a filtering function has been provided which only requires a subset of
the scheme details, e.g. the energy, renormalization scale and code, to be specified.
The following example demonstrates how to find the available process index tuples
and query them -- in this case for a spectrum file which contains cross-sections
at 8 TeV but none at 13 TeV:
%
\begin{Verbatim}[commandchars=\\\{\},frame=leftline,framesep=1.5ex,framerule=0.8pt,fontsize=\smaller]
\PY{o}{\PYZgt{}\PYZgt{}}\PY{o}{\PYZgt{}} \PY{n}{d} \PY{o}{=} \PY{n}{pyslha}\PY{o}{.}\PY{n}{read}\PY{p}{(}\PY{l+s}{\PYZdq{}}\PY{l+s}{myspectrum.slha}\PY{l+s}{\PYZdq{}}\PY{p}{)}
\PY{o}{\PYZgt{}\PYZgt{}}\PY{o}{\PYZgt{}} \PY{n}{d}\PY{o}{.}\PY{n}{xsections}\PY{o}{.}\PY{n}{keys}\PY{p}{(}\PY{p}{)}
\PY{p}{[}\PY{p}{(}\PY{l+m+mi}{2212}\PY{p}{,} \PY{l+m+mi}{2212}\PY{p}{,} \PY{l+m+mi}{1000001}\PY{p}{,} \PY{l+m+mi}{1000003}\PY{p}{)}\PY{p}{,}
 \PY{p}{(}\PY{l+m+mi}{2212}\PY{p}{,} \PY{l+m+mi}{2212}\PY{p}{,} \PY{o}{\PYZhy{}}\PY{l+m+mi}{1000002}\PY{p}{,} \PY{l+m+mi}{2000002}\PY{p}{)}\PY{p}{,} \PY{o}{.}\PY{o}{.}\PY{o}{.}\PY{p}{]}
\PY{o}{\PYZgt{}\PYZgt{}}\PY{o}{\PYZgt{}} \PY{n}{proc} \PY{o}{=} \PY{n}{d}\PY{o}{.}\PY{n}{xsections}\PY{p}{[}\PY{l+m+mi}{2212}\PY{p}{,} \PY{l+m+mi}{2212}\PY{p}{,} \PY{l+m+mi}{1000001}\PY{p}{,} \PY{l+m+mi}{1000003}\PY{p}{]}
\PY{o}{\PYZgt{}\PYZgt{}}\PY{o}{\PYZgt{}} \PY{n}{proc}\PY{o}{.}\PY{n}{get\PYZus{}xsecs}\PY{p}{(}\PY{n}{sqrts}\PY{o}{=}\PY{l+m+mf}{13000.}\PY{p}{,} \PY{n}{kappa\PYZus{}r}\PY{o}{=}\PY{l+m+mi}{2}\PY{p}{)}
\PY{p}{[}\PY{n}{sqrt}\PY{p}{(}\PY{n}{s}\PY{p}{)} \PY{o}{=} \PY{l+m+mi}{8000} \PY{n}{GeV}\PY{p}{,} \PY{n}{avg} \PY{n}{mass} \PY{n}{scale} \PY{n}{scheme}\PY{p}{,} \PY{o}{.}\PY{o}{.}\PY{o}{.}\PY{p}{]}
\PY{o}{\PYZgt{}\PYZgt{}}\PY{o}{\PYZgt{}} \PY{n}{proc}\PY{o}{.}\PY{n}{get\PYZus{}xsecs}\PY{p}{(}\PY{n}{sqrts}\PY{o}{=}\PY{l+m+mi}{13000}\PY{p}{)}
\PY{p}{[}\PY{p}{]}
\end{Verbatim}

\subsection{String representation}

\Block, \Particle and \Decay objects can all represent themselves in convenient
string form, as shown here for the NMIX matrix, neutralino LSP and gluino
entries read from an example mSUGRA file (and accessed via two \kbd{dict}{}s --
more on this later):
%
\begin{Verbatim}[commandchars=\\\{\},frame=leftline,framesep=1.5ex,framerule=0.8pt,fontsize=\smaller]
\PY{o}{\PYZgt{}\PYZgt{}}\PY{o}{\PYZgt{}} \PY{k}{print} \PY{n}{blocks}\PY{p}{[}\PY{l+s}{\PYZdq{}}\PY{l+s}{NMIX}\PY{l+s}{\PYZdq{}}\PY{p}{]}
\PY{n}{NMIX} \PY{p}{\PYZob{}} \PY{l+m+mi}{1}\PY{p}{,}\PY{l+m+mi}{1} \PY{p}{:} \PY{l+m+mf}{0.94321758}\PY{p}{;} \PY{l+m+mi}{1}\PY{p}{,}\PY{l+m+mi}{2} \PY{p}{:} \PY{o}{\PYZhy{}}\PY{l+m+mf}{0.20379469}\PY{p}{;} \PY{o}{.}\PY{o}{.}\PY{o}{.}
\PY{o}{\PYZgt{}\PYZgt{}}\PY{o}{\PYZgt{}} \PY{k}{print} \PY{n}{decays}\PY{p}{[}\PY{l+m+mi}{1000022}\PY{p}{]}
\PY{l+m+mi}{1000022} \PY{p}{:} \PY{n}{mass} \PY{o}{=} \PY{l+m+mf}{9.66880686e+01} \PY{n}{GeV} \PY{p}{:}
    \PY{n}{total} \PY{n}{width} \PY{o}{=} \PY{l+m+mf}{0.00000000e+00} \PY{n}{GeV}
\PY{o}{\PYZgt{}\PYZgt{}}\PY{o}{\PYZgt{}} \PY{k}{print} \PY{n}{decays}\PY{p}{[}\PY{l+m+mi}{1000021}\PY{p}{]}
\PY{l+m+mi}{1000021} \PY{p}{:} \PY{n}{mass} \PY{o}{=} \PY{l+m+mf}{6.07713704e+02} \PY{n}{GeV} \PY{p}{:}
    \PY{n}{total} \PY{n}{width} \PY{o}{=} \PY{l+m+mf}{5.50675438e+00} \PY{n}{GeV}
\PY{l+m+mf}{1.05840237e\PYZhy{}01} \PY{p}{[}\PY{l+m+mi}{1000005}\PY{p}{,} \PY{o}{\PYZhy{}}\PY{l+m+mi}{5}\PY{p}{]}
\PY{o}{.}\PY{o}{.}\PY{o}{.}
\end{Verbatim}


\subsection{File and string I/O}
\label{sec:io}

The objects previously described are the in-memory data which can be
programmatically manipulated by the user. To read these objects from SLHA files,
the functions \kbd{readSLHA()} and \kbd{readSLHAFile()} are supplied.

The format parsing itself lives in the former, which takes an SLHA file's
content as a string argument, and returns two \kbd{dict}s: one containing the
blocks keyed by name, and the the other containing the \Particle objects keyed
by PDG ID code\footnote{Again, these will be \emph{ordered} \kbd{dict}{}s if
  possible, ensuring that the order of blocks when iterated/written out matches
  that in which they were read in. The same applies to entries within
  blocks.}. The second (``\kbd{File}'') function treats its argument as either a
filename string or as a Python file object, from which it loads the file content
and passes it to \kbd{readSLHA()}. Both forms take an optional argument,
\kbd{ignorenobr}, which if set true will exclude any \Decay objects with a
branching ratio of 0 from the resulting \Particle{}s; it is set false by
default. Usage of these functions is demonstrated here:
%
\begin{Verbatim}[commandchars=\\\{\},frame=leftline,framesep=1.5ex,framerule=0.8pt,fontsize=\smaller]
\PY{n}{doc} \PY{o}{=} \PY{n}{pyslha}\PY{o}{.}\PY{n}{readSLHA}\PY{p}{(}\PY{n}{myslhastring}\PY{p}{)}
\PY{o}{.}\PY{o}{.}\PY{o}{.}
\PY{n}{doc} \PY{o}{=} \PY{n}{pyslha}\PY{o}{.}\PY{n}{readSLHAFile}\PY{p}{(}\PY{l+s}{\PYZdq{}}\PY{l+s}{sps1a.slha}\PY{l+s}{\PYZdq{}}\PY{p}{,}
                          \PY{n}{ignorenobr}\PY{o}{=}\PY{n+nb+bp}{True}\PY{p}{)}
\end{Verbatim}

SLHA format writing from code objects is similarly simple. Having placed \Block
and \Particle objects (the latter containing \Decay{}s) into \kbd{dict}{}s,
cf. the return values of the reader functions, they are passed to symmetric
writer functions as follows:
%
\begin{Verbatim}[commandchars=\\\{\},frame=leftline,framesep=1.5ex,framerule=0.8pt,fontsize=\smaller]
\PY{n}{slhastring} \PY{o}{=} \PY{n}{pyslha}\PY{o}{.}\PY{n}{writeSLHA}\PY{p}{(}\PY{n}{doc}\PY{p}{,}
                              \PY{n}{ignorenobr}\PY{o}{=}\PY{n+nb+bp}{True}\PY{p}{)}
\PY{o}{.}\PY{o}{.}\PY{o}{.}
\PY{n}{pyslha}\PY{o}{.}\PY{n}{writeSLHAFile}\PY{p}{(}\PY{l+s}{\PYZdq{}}\PY{l+s}{mymodel.slha}\PY{l+s}{\PYZdq{}}\PY{p}{,} \PY{n}{doc}\PY{p}{,}
                     \PY{n}{precision}\PY{o}{=}\PY{l+m+mi}{6}\PY{p}{)}
\end{Verbatim}

The optional \kbd{precision} argument specifies how many decimal places to be
written for floating point values such as masses, branching ratios, and widths:
the default is 8. Two further functions, \kbd{writeSLHABlocks(blo\-cks)} and
\kbd{writeSLHADecays(decays)} exist to separately produce the SLHA output
strings for the blocks and decays collections: both accept an optional
\kbd{precision} argument, and the optional \kbd{ignorenobr} argument may be
passed to the decay writer.


\subsection{HERWIG/ISAWIG $\leftrightarrow$ SLHA conversion}

An original major motivation for PySLHA was to convert SLHA spectrum/decay files
to the format used by the \Fortran HERWIG event
generator\,\cite{Corcella:2000bw,isawig} for SUSY simulation. This format was
previously only output by the ISAWIG program, which uses the SUSY spectrum
generator tools from the ISAJET code\,\cite{Paige:2003mg}.  The restriction of HERWIG
SUSY simulation to spectra generated by ISAJET/ISASUSY, and the increasing
difficulty of building ISAJET (due to reduced availability of CERNLIB and
Patchy, and increased standard enforcement in standard Fortran compilers)
motivated development of a format converter which would permit other spectrum
generators to produce HERWIG-compatible spectrum files.

Although use of ISASUSY (and of HERWIG) has reduced in recent years, PySLHA
retains the ability to read and write the ISAWIG format. The \kbd{readISAWIG()}
and \kbd{readISAWIGFile()} will parse the ISAWIG format into the PySLHA objects
cf. the \kbd{readSLHA*()} ones, and the \kbd{writeISAWIG()} and
\kbd{writeISAWIGFile()} functions invert the process. Unlike SLHA, in which the
data block format is arbitrarily extensible, the ISAWIG format is fixed and only
a subset of data will be written out: the corresponding block entries
\emph{must} be present.

There are some incompletenesses in PySLHA's ISA\-WIG support due to
a requirement that some SUSY decays be ordered in the file according to their
dependence on HERWIG internal matrix elements: it is not clear that these can be
handled in full detail without linking against HERWIG. Additionally, $R$-parity
violating couplings are not currently read in or written. These defects will be
happily resolved if possible and if there is sufficient interest.

Conversion between the ISAWIG and SLHA formats is made easier by two simple
converter scripts, \kbd{isawig2slha} and \kbd{slha2isawig}, so that programming
in Python is not necessary to convert a file of one sort into the other.

A side effect of having the ISAWIG $\leftrightarrow$ SLHA converter machinery in
PySLHA is that the library contains two functions, \kbd{herwigid2\-pdgid(hwid)}
and \kbd{pdgid2\-herwigid(pdgid)}, which as their names suggest convert particle
ID codes between the standard PDG scheme and the HERWIG internal scheme. This
may be useful, although undoubtedly less so as a new generation of C++ event
generator codes replace the venerable \Fortran ones.

\section{Mass spectrum and decay visualisation with \kbd{slhaplot}}

\begin{figure*}[t]
  \centering
  \includegraphics[width=0.85\textwidth]{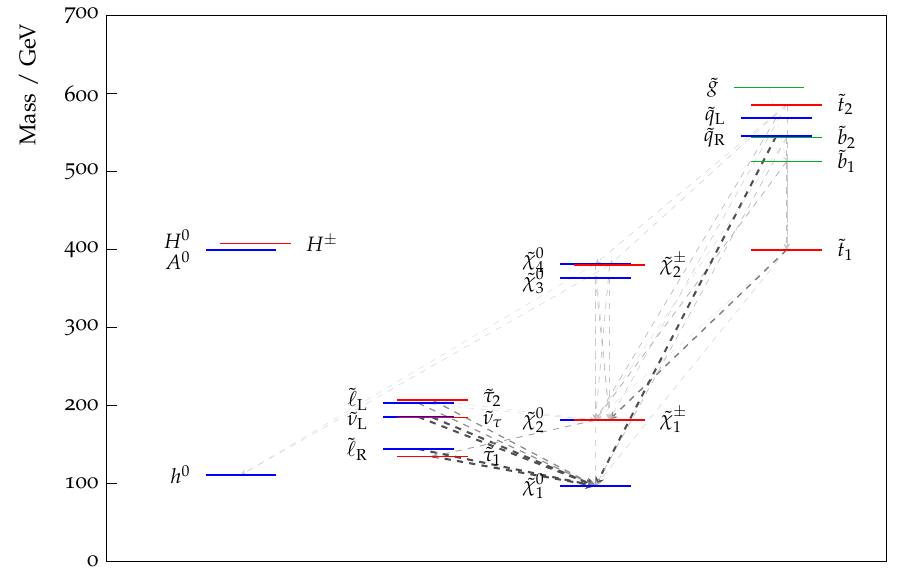}
  \caption{An example of the mass spectrum / decay channel visualisation output
    from \kbd{slhaplot}, showing the now-defunct SPS1 SUSY benchmark point.}
  \label{fig:slhaplot}
\end{figure*}

One of the most heavily used features of PySLHA, and a large part of the
original motivation for the package, is the \kbd{slhaplot} script, which
produces high quality graphical representations of SUSY mass spectra and decay
chains in a variety of formats including PDF, EPS, and \LaTeX. As usual, the
particle species are separated into distinct columns for Higgses, sleptons,
gauginos, and squarks/gluinos: an example is shown in Figure~\ref{fig:slhaplot}.

Use of \kbd{slhaplot} is simple: for default output a number of SLHA (or ISAWIG)
filenames are supplied on the command line and each will produce a resulting PDF
plot (or plots) with the same base name. A \kbd{format} flag may be given to
specify which output formats to use, as a comma-separated list: more than one is
allowed if e.g. output of the same plot in PDF, EPS, PNG and \LaTeX{} formats
were to be useful. Fundamentally the plots are produced using the
PGF/TikZ\,\cite{tikz} \TeX-based graphics library, and the \kbd{tex} output format
contains the raw TikZ programming statements for modification by the end user if
wanted. A \LaTeX{} preamble may be supplied to \kbd{slhaplot} by users who wish
to e.g. use different fonts, and another option allows a TikZ fragment to be
included for easy adding of drawing items, e.g. extra labels on the plot.

Many further rendering options are available via the \kbd{slhaplot} command
line: these include
\begin{itemize}
\item a minimum branching ratio to be displayed, to avoid cluttering the plots
  with virtually non-existent decays;
\item a maximum particle mass to be displayed; by default this is calculated
  from the mass spectrum, and the plot axis extent chosen for cosmetic
  appropriateness;
\item variations on the rendering style for decay arrows (BR-dependent line
  widths and line colours);
\item variations in mass-line labelling algorithms -- \kbd{slhaplot} makes
  efforts to avoid ugly and unhelpful overlaps of particle name labels resulting
  from near-degenerate masses. The user may choose either to merge close-by
  labels into a single label, or to shift labels by small amounts to avoid the
  clashes. The latter is the default behaviour and produces good results most of
  the time: if fine control is needed, the user may dump and modify the \LaTeX{}
  plot source code.
\item changing the plot aspect ratio.
\end{itemize}

The Python rendering engine which calls \kbd{latex},\linebreak[4] \kbd{pdflatex}, and format
converter programs (with caching of intermediate stages for efficiency) has been
split off from PySLHA as the separate \kbd{tex2pix} package\,\cite{tex2pix}.

\section{Summary}

PySLHA is a mature Python library for reading, writing and manipulating SUSY
model data with input and output to the SLHA and ISAWIG formats. Also included
are scripts for easy conversion between these two data formats, and for
producing publication-ready SUSY mass-spectrum/decay plots.

PySLHA is available from the Python package index (PyPI) at
\url{https://pypi.python.org/pypi/pyslha} and a home page with more usage
instructions is available at \url{http://insectnation.org/projects/pyslha}. The
code is documented according to normal Python standards which may be queried via
\kbd{pydoc} or the Python interactive \kbd{help()} command. The latest version
at the time of writing is PySLHA 3.1.1.

\begin{acknowledgements}

I started PySLHA for personal amusement and challenge (and with the hope of
avoiding future compilations of ISAJET/CERNLIB) at an MCnet summer school in
2010. It was later extended -- in particular with the addition of \kbd{slhaplot}
-- for use in graduate lectures on BSM physics organised by the Scottish
Universities Physics Alliance (SUPA). My thanks both to MCnet and SUPA for their
support via travel funding and a research fellowship respectively.

I further acknowledge and thank the Royal~Society for their funding of a
University Research Fellowship, CERN for a visiting Scientific Associateship,
and the Durham~University Institute for Particle Physics Phenomenology for
several small associateship grants: all these have greatly assisted with
collaborative work on particle physics MC modelling, PySLHA included.

Thanks also to the many users who made development suggestions (such as writing
this paper!) and let me know that PySLHA had active users; without their
appreciation and enthusiasm it would have been all too easy to let the project
stagnate once it met my own needs.

\end{acknowledgements}


\end{document}